# Effective gap at microwave frequencies in $MgB_2$ thin films with strong interband scattering

G. Ghigo, D. Botta, A. Chiodoni, L. Gozzelino, R. Gerbaldo, F. Laviano, and E. Mezzetti

Department of Physics, Politecnico di Torino, C.so Duca degli Abruzzi 24, 10129 Torino, Italy,

Istituto Nazionale di Fisica Nucleare, Sez. Torino, via P. Giuria 1, 10125 Torino, Italy

and

Istituto Nazionale per la Fisica della Materia, U.d.R. Torino-Politecnico, C.so Duca Degli Abruzzi 24, 10129 Torino, Italy.

E. Monticone and C. Portesi

Istituto Elettrotecnico Nazionale G.Ferraris, strada delle Cacce 91, 10135 Torino, Italy

The microwave properties of polycrystalline $MgB_2$ thin films prepared by the so-called in-situ method are investigated. The characterization of the films at microwave frequencies was obtained by a coplanar resonator technique. The analysis of the experimental data results in the determination of penetration depth, surface impedance and complex conductivity. The aim of this work is to set the experimental results in a consistent framework, involving the two-band model in the presence of impurity scattering. The energy gaps are calculated and the contribution of intra- and inter-band scattering is considered. From the comparison between the calculated gap values and the experimental data it turns out that the temperature dependence of the penetration depth can be accounted for by an effective mean energy gap, in agreement with the predictions of Kogan et al. [Phys. Rev. B **69**, 132506 (2004)]. On the other hand, the temperature dependence of the real part of the microwave conductivity and of the surface resistance is accounted for by the single smaller gap, in agreement with the work of Jin et al. [Phys. Rev. Lett. **91**, 127006 (2003)]. Since these findings rely on the same calculated gap structure, the required consistency is fulfilled.

## Introduction

Since the discovery of its superconducting properties, magnesium diboride ($MgB_2$) has generated a great deal of interest because of its simple structure, relatively high critical temperature and two-gap nature. The Fermi surface of $MgB_2$ consists of two three-dimensional sheets, from the $\pi$ bonding and antibonding bands, and two nearly cilindrical sheets from the two-dimensional $\sigma$ bands.[1] Many physical properties of $MgB_2$ are reasonably described within a model with two separated energy gaps, $\Delta_\pi$ and $\Delta_\sigma$.[2,3] Nevertheless, the role of interband and intraband scattering has to be considered:[2,3] it is still not completely clear, also due to the wide spread quality of samples used in different experiments. In fact, a significant scattering between the different Fermi sheets may reduce the effective gap structure to a single isotropic gap. Recently, the expected observation of single-gap superconductivity at high impurity level has been observed in C-substituted $MgB_2$ single crystals by point-contact spectroscopy.[4] When the level of impurities is high enough, the two gaps merge into a single gap with a ratio $2\Delta/k_BT_c$ close to the standard BCS value.

The role of the two bands in determining the microwave conductivity in $MgB_2$ thin films is controversial as well. The temperature dependence of the microwave conductivity in c-axis oriented films was interpreted by Jin et al. [5] in terms of a dominant contribution of the $\pi$ band. They deduced this argument from the observation of a single anomalous peak around $t=T/T_c=0.6$. The presence of such a peak can be explained in the framework of the BCS theory. When the superconducting state is entered and the gap opens, a singularity appears in the density of states at the gap edges, increasing the microwave conductivity.[6] When the gap becomes larger than $k_BT$, upon lowering the temperature, quasiparticles condense and microwave conductivity is suppressed. The peak at $t=0.6$ is consistent with a small gap. On the other hand, a clear statement in favor of the contribution of both the bands in determining the microwave conductivity in their high quality polycrystalline films ($T_c$=39K) comes from Lee et al.[7] They observed two distinct peaks in the conductivity, at $t=0.5$ and $t=0.9$, consistent with the expected $\Delta_\pi$ and $\Delta_\sigma$ values.

It is worth mentioning that the properties connected to superconductivity and to electric transport do not necessarily rely on the same band, in accordance with the arguments reported in ref.8. As a consequence, a comparison between different properties measured on the same sample, such as penetration length, surface resistance or complex microwave conductivity, and the supposed band structure is needed to shed light on the issue. This is in fact the aim of this work, where the microwave properties of $MgB_2$ thin films prepared by the so-called in-situ method are investigated. The characterization of the films at microwave frequencies, obtained by a coplanar resonator technique, is presented. From the analysis of the experimental data we determine the penetration depth, the surface impedance and the complex conductivity. The relatively low $T_c$ value (about 30K) suggests the presence of a relatively high level of impurities, which enhance the interband scattering. It is worthwhile to note that also the ion milling process, needed to obtain the resonators, increases the disorder in the pristine film. Accordingly, we consider the contribution of intra- and inter-band scattering, in the framework of the model proposed by Kogan et al., in order to obtain a reliable description of the temperature dependence of the measured penetration depth. The resulting gap structure is also compared to the results concerning the complex microwave conductivity and is discussed in detail.

## Preparation and microwave measurements

We fabricated 1cm×1cm×110nm $MgB_2$ thin films on (0001) sapphire substrates by a co-evaporation technique followed by in-situ annealing. B was evaporated by e-gun and Mg was evaporated by a resistive heater. During deposition, the substrate was held at a temperature of 280 ºC. The thermal treatment was carried out at 500 ºC for 5 minutes. Using this method, we obtained $MgB_2$ thin films with rather good electrical properties and smooth and homogeneous surfaces. The analysis of the morphological and structural properties has been performed by means of AFM and XRD, respectively. The roughness of the samples ranges between 10 and 20 nm (if evaluated on 5 $\mu m^2$ areas), and depends on the film thickness. The films are polycrystalline and in the best samples we observe a partial orientation along the c-axis.

Linear coplanar resonators have been obtained by a standard photolithographic process followed by dry etching in an ion milling system: we obtained well-defined structures with very sharp edges. The length of the central conductor is $l$=8mm, its width is $w$=300μm and the distance between the ground planes is $a$=700μm.

The microwave device was cooled in an Oxford He-flow cryostat equipped with an ITC503 temperature controller and with a custom-made dc magnet with high field uniformity. By means of Rohde-Schwarz ZVK vector network analyzer we measured the complex transmission coefficient, $S_{21}$ (ratio of the voltage transmitted to the incident voltage [9]), as a function of the driving frequency, $f$. Fig.1 shows the resonance curves $S_{21}(f)$ at increasing temperatures (inset) and at T=4.52K (main panel). The model of the resonator as a RLC circuit, needed to extract by means of a fit of the experimental data the relevant parameters (resonant frequency $f_0$, loaded quality factor $Q_L$ and unloaded quality factor $Q_0$) has been reported elsewhere.[10] The best fit for the resonant curve at $T$=4.52K is shown in fig.1 (solid line).

The resonator results to be quite sensitive to rf and dc magnetic fields, both directed perpendicularly to the film surface. Fig.2 shows the decrement of the loaded quality factor, $Q_L$, as a function of external dc field and input power, $P_{in}$, at $T$=5K. This high sensitivity can be attributed to the high aspect ratio (thin film in a transverse field), which causes magnetic field line focusing at the edges, and to the coplanar layout, which induces rf current peaks at the edges, and finally to the intrinsic properties of $MgB_2$. In the following sections we analyze only microwave data measured in zero dc magnetic field and $P_{in}$=–20dBm, since the upper limit of the linearity zone is about –10 dBm. Fig.3 shows the resonant frequency and the unloaded quality factor as a function of temperature. These parameters represent the final output of the measurement procedure and are the base for the following analysis, aimed at determining all the physical quantities.

## Determination of penetration depth, surface resistance and microwave complex conductivity

The analysis of the microwave data, aimed at the evaluation of the penetration depth and the surface resistance, has to suitably account also for the substrate properties.[11] In the standard theory of distributed element transmission lines for a half-wavelength resonator,[12] the resonant frequency is given by

$$f_0 = \frac{1}{2l\sqrt{L_l C_l}} \qquad (1)$$

where $l$ is the length of the resonator, $L_l$ and $C_l$ are the inductance and the capacitance per unit length, respectively. The ratio

$$\frac{f_0(T)}{f_0(T_0)} = \sqrt{\frac{L_l^g + L_l^k(\lambda(T_0))}{L_l^g + L_l^k(\lambda(T))} \frac{C_l(\varepsilon_r(T_0))}{C_l(\varepsilon_r(T))}},$$

where $L_l^g$ and $L_l^k$ are the geometrical and kinetic inductances respectively, depends on temperature through the penetration depth $\lambda(T)$ and the permittivity of the substrate $\varepsilon_r(T)$. $T_0$ is a reference temperature, usually the lowest in the measured range. We fit the $f_0(T)/f_0(T_0)$ experimental data with suitable parametric expressions for the temperature dependence of $\lambda$ and $\varepsilon_r$, and with standard formulas for coplanar waveguides,[11,13] as shown in fig.4. The temperature dependence of the London penetration depth here assumed in the fitting procedure is

$$\lambda_L(T) = \frac{\lambda_L(0)}{\sqrt{1-(T/T_c)^\gamma}} \qquad (2)$$

with $\gamma = 3\text{-}T/T_c$, describing the weak coupling regime.[14] Attempts to use $\gamma = 4$ (strong coupling regime and two fluid approximation) or $\gamma = 2$ (d-wave superconductivity) [15] did not lead to any reasonable result. It has to be noted that (2) applies only to homogeneous superconductors with a single isotropic gap. The fact that this expression fits very well the experimental data is a first indication that our polycrystalline $MgB_2$ films have significant scattering between the different Fermi sheets, which may reduce the effective gap structure to a single isotropic gap. Significant deviations from (2) are expected close to $T_c$, where the microwave penetration depth $\lambda$ deviates from the London penetration depth $\lambda_L$ in order to meet the normal skin depth. Therefore the fit was performed in a suitable temperature range, where we estimate that the measured penetration depth $\lambda \approx \lambda_L$ (i.e. where $\lambda_L^{-2} >> \omega\mu_0\sigma_1$, being $\sigma_1$ the real part of the conductivity). This procedure allows obtaining a reliable evaluation of $C_l(T)$, that we can extrapolate to the whole measured temperature range. Than we can recalculate $L_l(T)$ from (1) by means of $C_l(T)$ and of the experimental data $f_0(T)$. From these new $L_l(T)$ values, extended to the whole temperature range, we extract the final λ($T$) values by means of the formulas for coplanar lines.[11,13] This procedure assures a model-independent determination of λ (inset of fig.4). The value λ(0)≈260nm deduced from the data fitting seems to be quite large if compared with other estimates in literature. It can be attributed to the fact that the film contains a certain level of impurities, as the relatively low critical temperature shows: impurities and interaction effects drastically enhance the penetration depth. In fact this value is in between the estimates $\lambda_{L,ab}^{dirty}$=105.7nm and $\lambda_{L,c}^{dirty}$=316.5nm given by Golubov et al. A similar result is reported in ref.15 (film I) where data show $T_c$=29K, λ(0)=300nm and are fitted by (2) with γ = 3-$T/T_c$ as well.

The surface resistance, $R_s$, is deduced as

$$R_s = 2\pi \frac{f_0 L(\lambda) w_{eff}(\lambda)}{Q_0}$$

where $w_{eff}$ is the effective width of the line and $\lambda = \lambda_L \coth(d/2\lambda_L)$. The dependence of the surface resistance on temperature is shown in fig.5. Despite the relatively low $T_c$, the film shows a quite low $R_{res}$, comparable with values in high-quality films:[16,17] $R_{res}$=10 μΩ if the low temperature points are extrapolated to $T$=0K, as shown in fig.5, or $R_{res}$=32 μΩ as deduced by the fit shown below (fig.9).

The real and imaginary parts of the complex microwave conductivity, related to the complex surface impedance through $Z_s = R_s + iX_s = \sqrt{i\mu_0\omega/(\sigma_1 - i\sigma_2)}$, can be obtained as [18]

$$\sigma_1 = \frac{2\mu_0\omega R_s X_s}{(2R_s X_s)^2 + (R_s^{\,2} - X_s^{\,2})^2} \quad ; \quad \sigma_2 = \frac{\mu_0\omega (X_s^{\,2} - R_s^{\,2})}{(2R_s X_s)^2 + (R_s^{\,2} - X_s^{\,2})^2}$$

where $X_s = 2\pi f_0 \mu_0 \lambda^2$ , and are shown in fig.6.

## Effective gap

The physical parameters obtained above should somehow mirror the gap structure of the material. In this section we determine the gap functions underlying the experimental data from the comparison between measurements and calculations based on two-gap models. In order to proceed, we preliminarily examine the temperature dependence of the penetration depth. For a clean system, a kink or at least an inflection point in the temperature dependence of the penetration depth is predicted in the range of $T/T_c$ from 0.3 to 0.5.[1,19] Interband scattering is expected to smooth out this feature [1,19] and its absence in our data supports the hypothesis that interband effects are rather strong. Also the relatively low critical temperature is a further indication of the importance of interband impurity scattering. Accordingly, we consider the following expression for the London penetration depth in the case of strong interband and intraband scattering, as given by Kogan et al.:

$$\left(\lambda^2\right)_{ik}^{-1} \propto \Delta^* \tanh\frac{\Delta^*}{2k_B T}\sum_\alpha \nu_\alpha \left\langle v_i v_k \right\rangle_\alpha \tau_\alpha \ , \qquad (3)$$

where $\alpha=\pi,\sigma$, $\nu_\alpha$ are relative densities of states, $v_\alpha$ are the Fermi velocities, $\tau_\alpha$ are suitable combinations of scattering times and $\Delta^*$ is an effective mean energy gap

$$\Delta^* = \frac{(\zeta_\pi + \zeta_\sigma)\Delta_\pi \Delta_\sigma}{\zeta_\pi \Delta_\pi + \zeta_\sigma \Delta_\sigma} \ ; \ \zeta_\pi = \frac{\nu_\sigma}{2\tau_{\pi\sigma}\Delta_\pi} \ ; \ \zeta_\sigma = \frac{\nu_\pi}{2\tau_{\sigma\pi}\Delta_\sigma} \qquad (4).$$

$\zeta_\pi$ and $\zeta_\sigma$ are proportional to the interband scattering level ($\zeta_\pi,\zeta_\sigma>>1$ for strong interband scattering ). If the terms in the sum in (3) are temperature independent in the investigated range, the following expression holds:

$$\frac{\lambda^2(0)}{\lambda^2(T)} = \frac{\Delta^*(T)}{\Delta^*(0)} \tanh \frac{\Delta^*}{2k_B T} \qquad (5)$$

A comparison between the measured $\lambda(T)$ and calculations based on (5) are shown in fig.7, where different theoretical curves are reported and main parameters are listed. The calculation of the two gaps $\Delta_\pi$ and $\Delta_\sigma$, needed to obtain $\Delta^*$ through (4), is performed in the framework of the two-band Eliashberg theory and proceeds as follows.[20] The critical temperature of an ideal impurity-free $MgB_2$ film, $T_c$*, is fixed by a suitable $\mu$ value (prefactor of the Coulomb pseudopotential). The effect of impurities is then added by setting the interband scattering rate $\Gamma_{\pi\sigma}$, see Ref.20, to a value adjusted to get the experimental $T_c$, starting from $T_c$*. We assume the spectral functions and the electron-phonon coupling constants calculated by Golubov et al.[21,22] ($\lambda_{\pi\pi}$=0.448, $\lambda_{\sigma\sigma}$=1.017, $\lambda_{\sigma\pi}$=0.213, $\lambda_{\pi\sigma}$=0.155), the Coulomb pseudopotential as in Ref.20, with a cutoff energy of 700 meV and the solution calculated until a maximum energy of 800 meV. The considered densities of states at the Fermi level in the $\sigma$ and $\pi$ band are 0.3 and 0.4 states/(eV unit cell), respectively. Once obtained $\Delta_\pi$ ($T$) and $\Delta_\sigma$ ($T$), the penetration depth is deduced through (4) and (5), with $\nu_\pi$ =0.43, $\nu_\sigma$ =0.57 and scattering times $\tau_{\pi\sigma} = \tau_{\sigma\pi}$ =1/ $\Gamma_{\pi\sigma}$, and compared to experimental data. All the curves reported in fig.7 are calculated by using $\mu$ and $\Gamma_{\pi\sigma}$ values that give the same $T_c$, the experimental one. The best agreement between theoretical and experimental curves is found in the case $\mu$=0.0475 and $\Gamma_{\pi\sigma}$ =0.8meV (curve #1; the corresponding gap functions are reported in fig.8.). The resulting value of $\Gamma_{\pi\sigma}$ corresponds to the case of intermediate-to-strong interband scattering, but fulfill the condition $\Gamma_{\pi\pi}>\Gamma_{\sigma\sigma}>>\Gamma_{\pi\sigma}$ suggested in ref.8. In this case it should be noted that if the superconducting band with the smaller gap is overdamped due to impurities, then the penetration depth is expected to be dominated by the other band. In fact, we notice that the $\lambda(T)$ dependence in our films is much closer to the curve expected for pure $\sigma$ contribution than to the curve expected for pure $\pi$ contribution (fig.7, curves #4 and #5, respectively).

An opposite behavior, i.e. a dominant role of the smaller gap, is expected when the properties connected to dissipation and electric transport are concerned, in agreement with the arguments of ref.8. We now discuss the temperature dependence of the real part of the microwave conductivity, $\sigma_1$, and of the surface resistance, $R_s$, by comparing data to the gap structure we obtained above (fig.8). The aim is twofold. First of all, we look for further support to the derivation of the gaps from the penetration depth. Moreover, we want to check if the prevailing influence of the smaller gap, claimed for $\sigma_1$ and $R_s$, holds.

The dotted line in fig.8 represents the condition $\Delta(T)=k_BT$. This line crosses the gap curves in correspondence of reduced temperatures that could be relevant to discuss the dependence on temperature of $\sigma_1$. As already mentioned in the Introduction, the peak shown by $\sigma_1(T)$ is connected to the value of the gap. In our case, the $\sigma_1$ peak occurs at t≈0.62 (fig.6), exactly at the same temperature where the $\Delta(T)=k_BT$ line crosses the $\Delta_\pi(T)$ curve in fig.8. This result represents a remarkable agreement between independent physical parameters framed into a consistent two-gap model. It turns out that $\sigma_1$ cannot decrease until quasiparticles start to condensate also in the $\pi$ band, in accordance with the findings of Jin et al. No clear features in the $\sigma_1(T)$ curve emerge at reduced temperatures corresponding to the intersections of the $\Delta(T)=k_BT$ line with either $\Delta_\sigma(T)$ or $\Delta^*(T)$ functions.

Since the behavior of $\sigma_1(T)$ is ruled by the smaller gap and $\lambda(T)$ results from a mean effective gap, the question arises about the prevailing effect on $R_s$. In fact, in a two-fluid model, $R_s \cong \left(\omega^2 \mu_0^2 \lambda^3 \sigma_1\right)/2$. Low temperature $R_s$ data can be used to estimate the gap value according to the fitting expression [23]

$$R_s(t) = \frac{A}{t} \ln\left(\frac{4t}{\nu\delta}\right) e^{-\delta/t} + R_{res} \qquad (6)$$

where $t=T/T_c$, $\nu=hf/\Delta$, $\delta = \Delta(0)/kT_c$ and $R_{res}$ is the residual resistance. Fig.9 shows that the experimental data can be nicely fitted by (6), with $\Delta(0)$ fixed to $\Delta_\pi(0)$ and with only three free parameters left (solid line). Attempts to get a good fit by setting $\Delta(0)$ to $\Delta_\sigma(0)$ or to $\Delta^*(0)$ did not lead to reasonable results (dashed and dotted lines, respectively). This result rules out the contribution of the $\sigma$ band to $R_s$, in the investigated temperature range. Its validity has to be checked for $T>T_c/2$, since the cubic dependence of $R_s$ on $\lambda$ is expected to prevail.

## Conclusions

The microwave properties of polycrystalline $MgB_2$ thin films prepared by the in-situ method have been investigated by a coplanar resonator technique. Penetration depth, surface resistance and microwave conductivity were extracted by the analysis of the resonance curves at different temperatures. Data were interpreted by considering the contribution of intra- and inter-band scattering, in the framework of the two-gap model proposed by Kogan et al. It allows obtaining an effective energy gap from the two gap functions, which have been calculated from the two-gap Eliashberg theory. From the comparison between the calculated gap values and the experimental data it turns out that the temperature dependence of the penetration depth can be accounted for by the effective mean energy gap. On the other hand, the temperature dependence of the real part of the microwave conductivity and of the surface resistance is accounted for by the single small $\pi$ gap, in agreement with other literature findings. These results are fully consistent since they rely on the same calculated gap structure. The main microwave properties of the investigated films make them promising for applications such as kinetic inductance photon detectors.

## Acknowledgments

The authors whish to thank G. A. Ummarino (Politecnico di Torino, I) for the solution of the two-band Eliashberg equations and F. Roesthuis (University of Twente, NL) for his help in the ion milling process. This work has been partially supported by INFN and ASI under the project CAME200310 and by the Italian MIUR under the FIRB-RBAU01PEMR project.

**Figure 1. Resonance curve at $T$=4.52K. The solid line represents the best fit of the experimental data (see text). Inset: resonance curves at increasing temperatures.**

**Figure 2. Percent decrement of the loaded quality factor, $Q_L$, as a function of the input power, $P_{in}$, at zero dc field (solid symbols) and as a function of the external dc field at $P_{in}$=-20dBm (open symbols).**

**Figure 3. Resonant frequency (right) and unloaded quality factor (left) as a function of temperature.**

**Figure 4. Normalized resonant frequency as a function of temperature. The solid line is the best fit of the experimental data (symbols) obtained by the procedure described in the text. The inset shows the penetration depth as deduced from the fit (London penetration depth, solid line) and as recalculated from (1) (microwave penetration depth, symbols, see ref.11).**

**Figure 5. Surface resistance as a function of temperature. The dotted line is the extrapolation of low temperature data to $T$=0K. Inset: the same data, in a linear scale.**

**Figure 6. Real part, $\sigma_1$, and imaginary part, $\sigma_2$, of the microwave complex conductivity as a function of temperature. Note the different scales concerning the real (right) and imaginary parts (left).**

**Figure 7. Temperature dependence of the penetration depth: experimental (symbols) and theoretical curves (lines). The theoretical curves are calculated following the procedure described in the text, with the reported parameters.**

**Figure 8. Gap structure, as determined by the comparison between experimental data and theoretical curves in fig.7 (see also the text). The arrow marks the reduced temperature at which $\Delta_\pi=k_B T$.**

**Figure 9. Low temperature surface resistance data, fitted by (6) with the parameter $\Delta(0)$ set to $\Delta_\pi(0)$, $\Delta_\sigma(0)$ and $\Delta^*(0)$, respectively. The residual resistance, $R_{res}$, resulting from the different fits is also reported.**

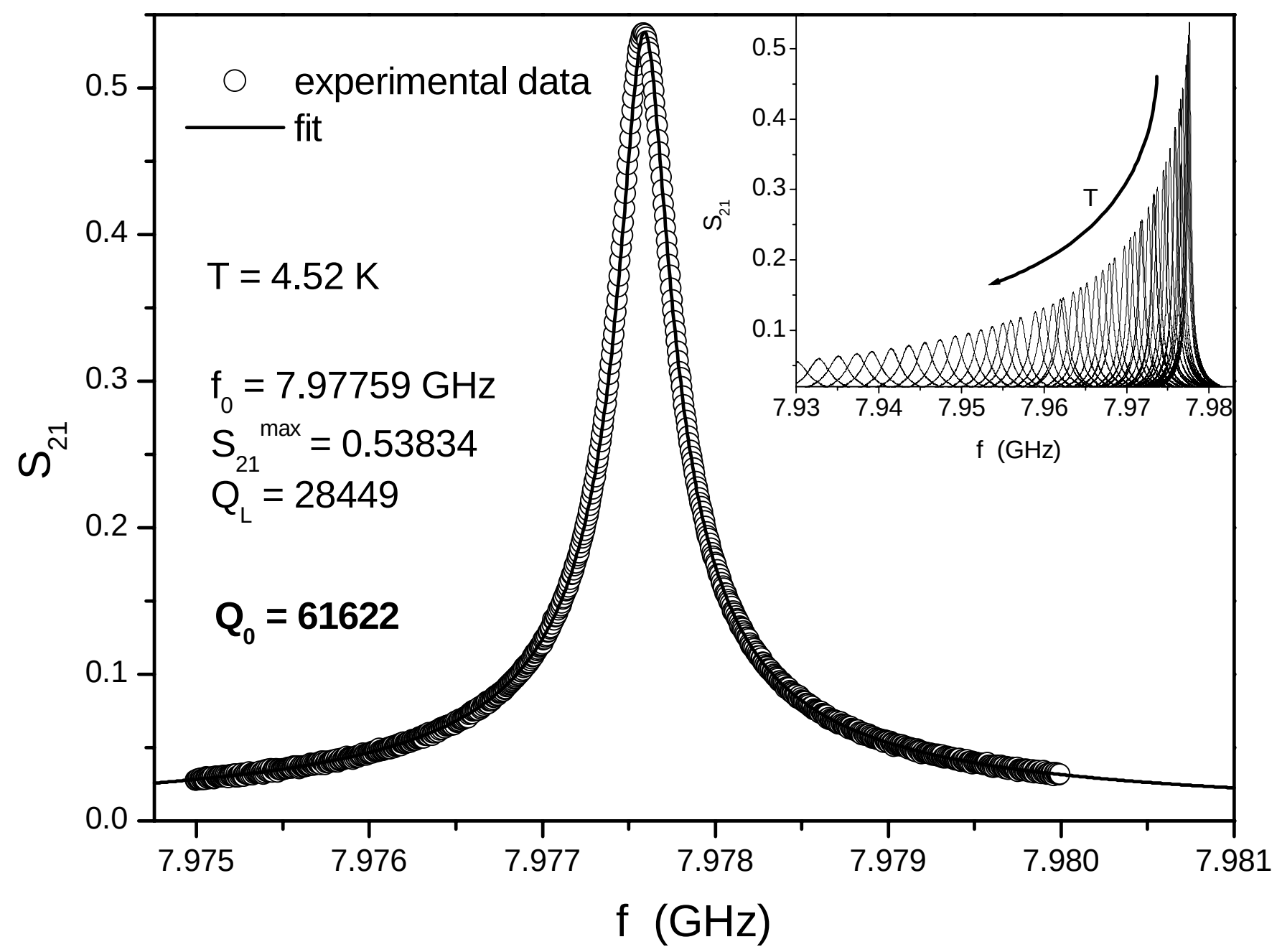

**Figure 1**

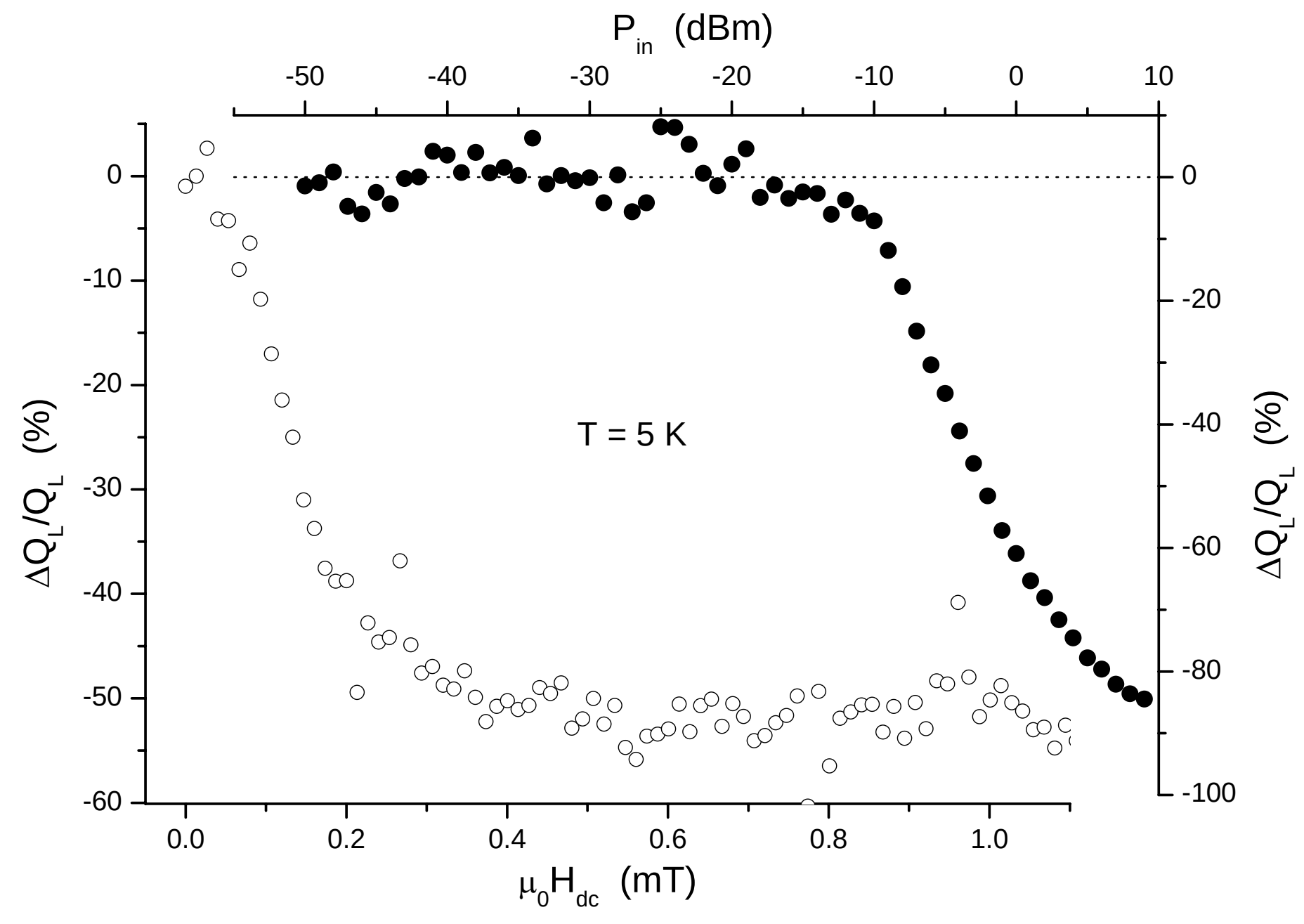

**Figure 2**

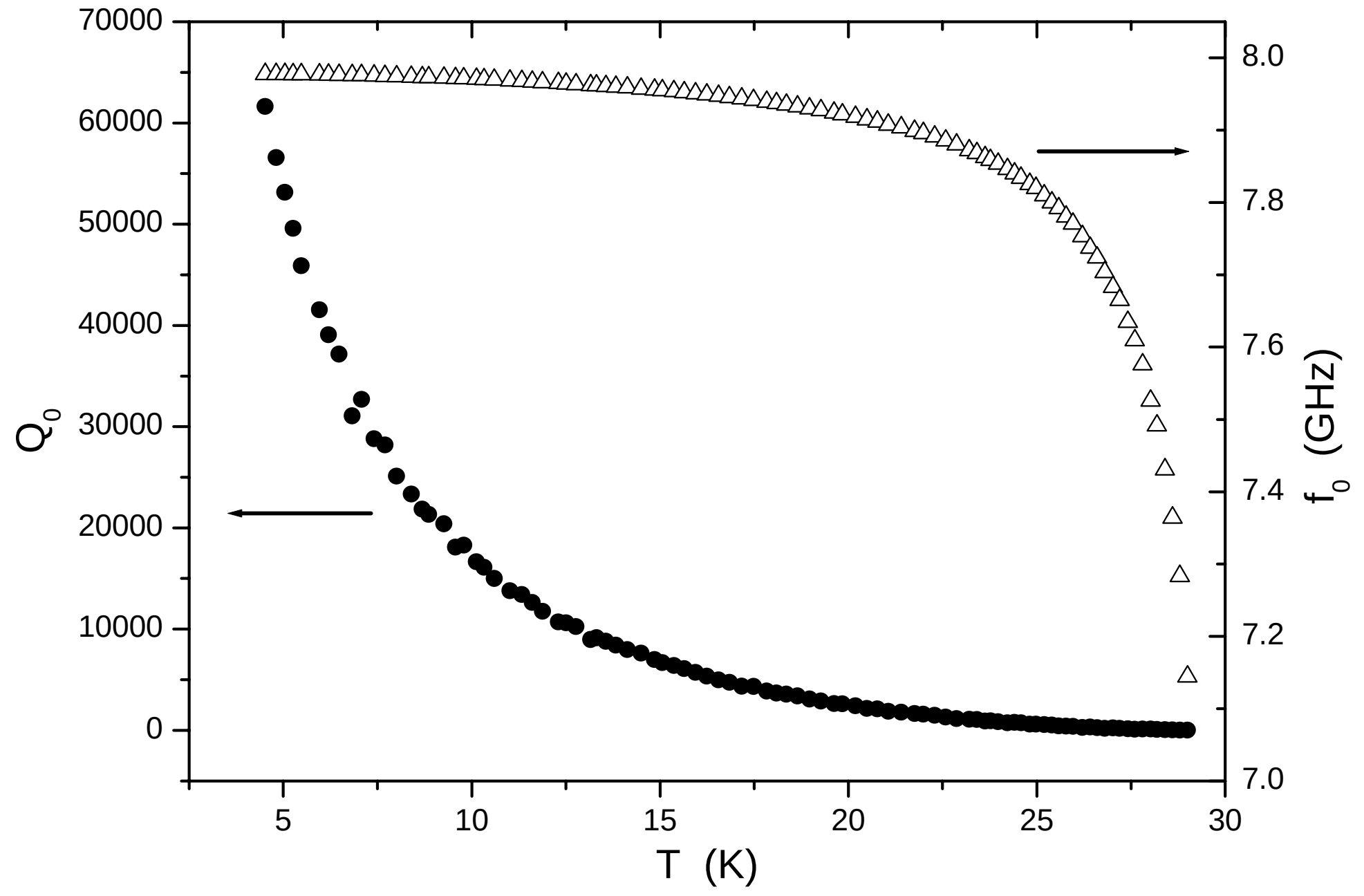

**Figure 3**

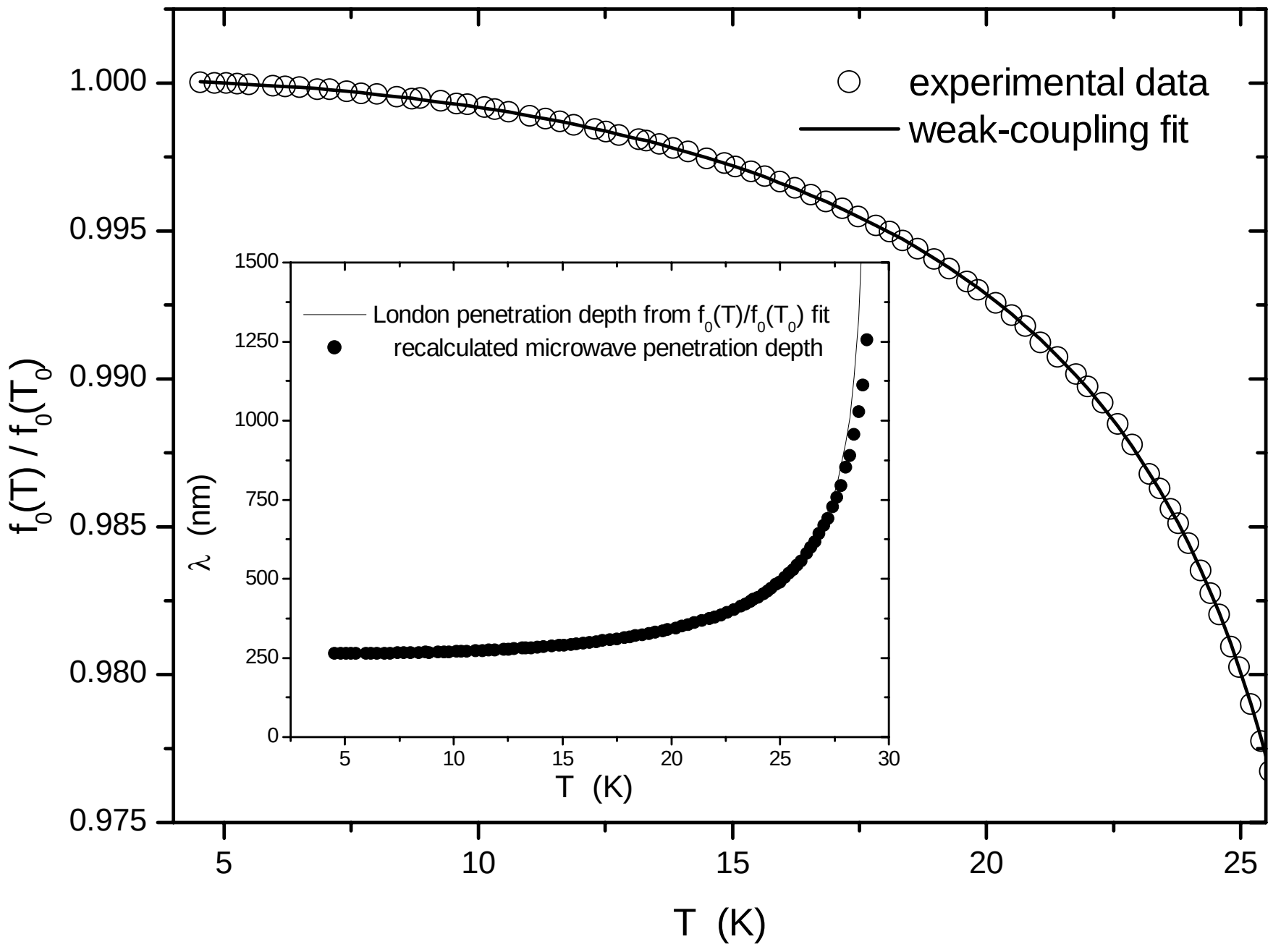

**Figure 4**

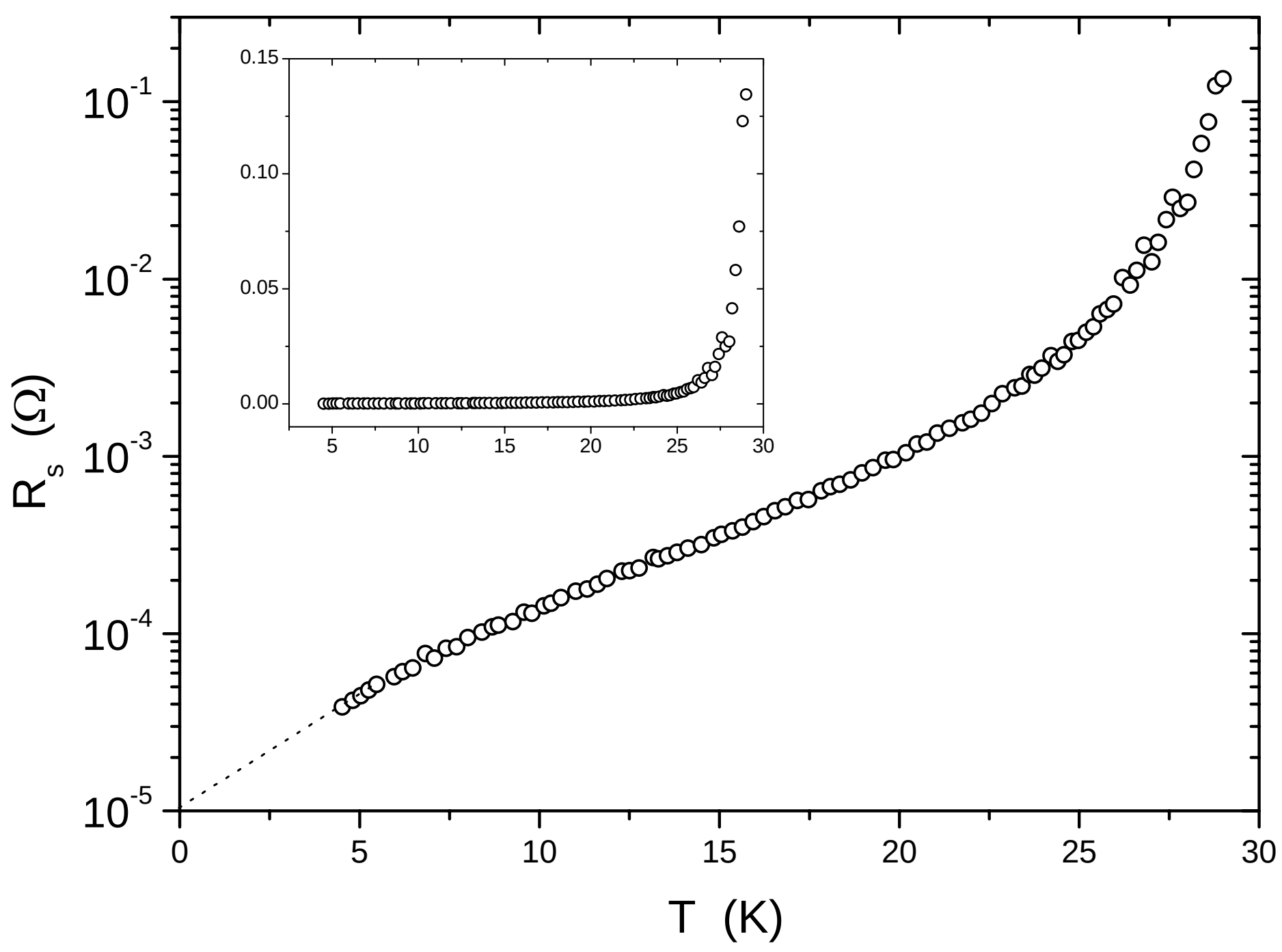

**Figure 5**

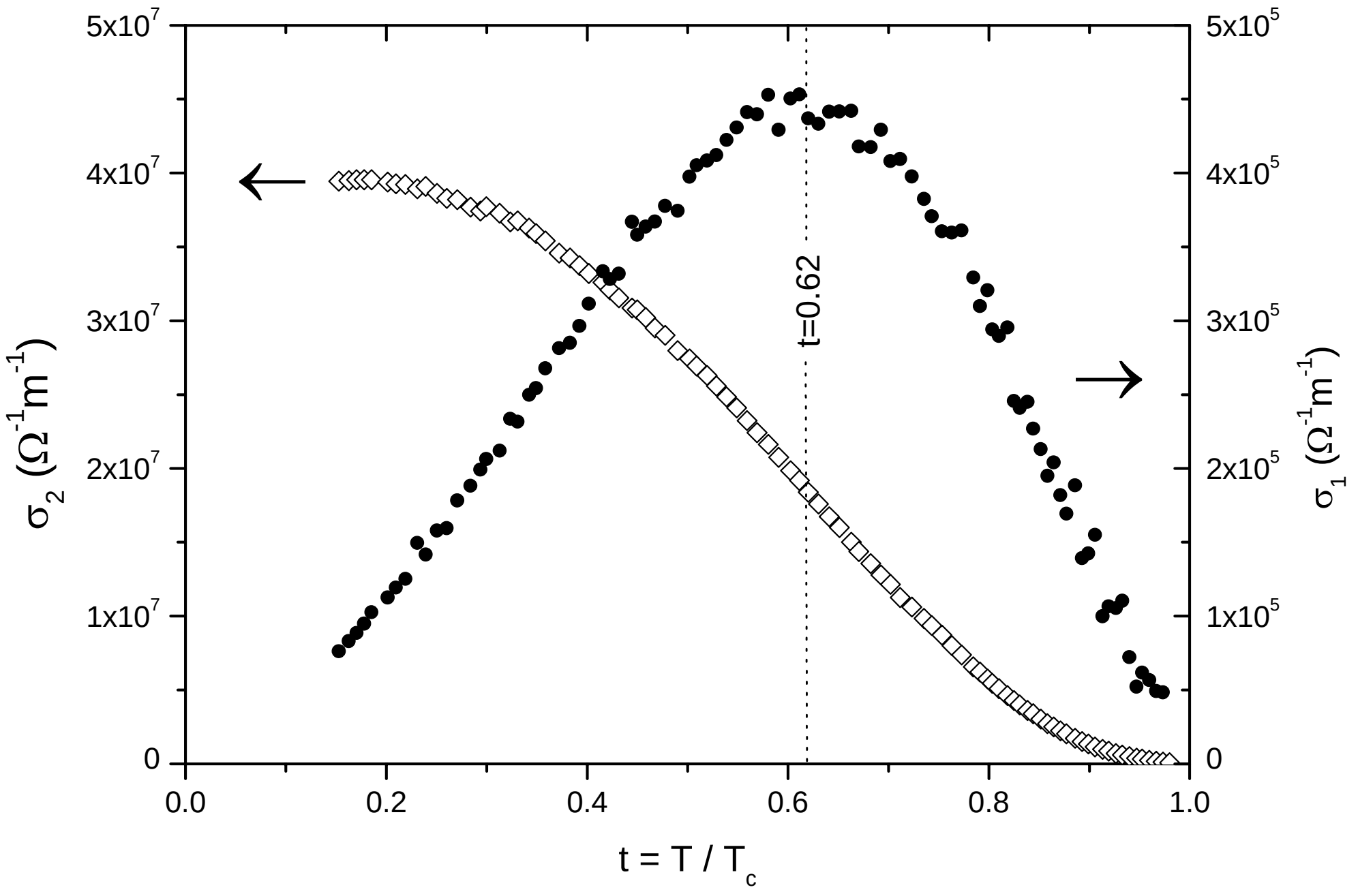

**Figure 6**

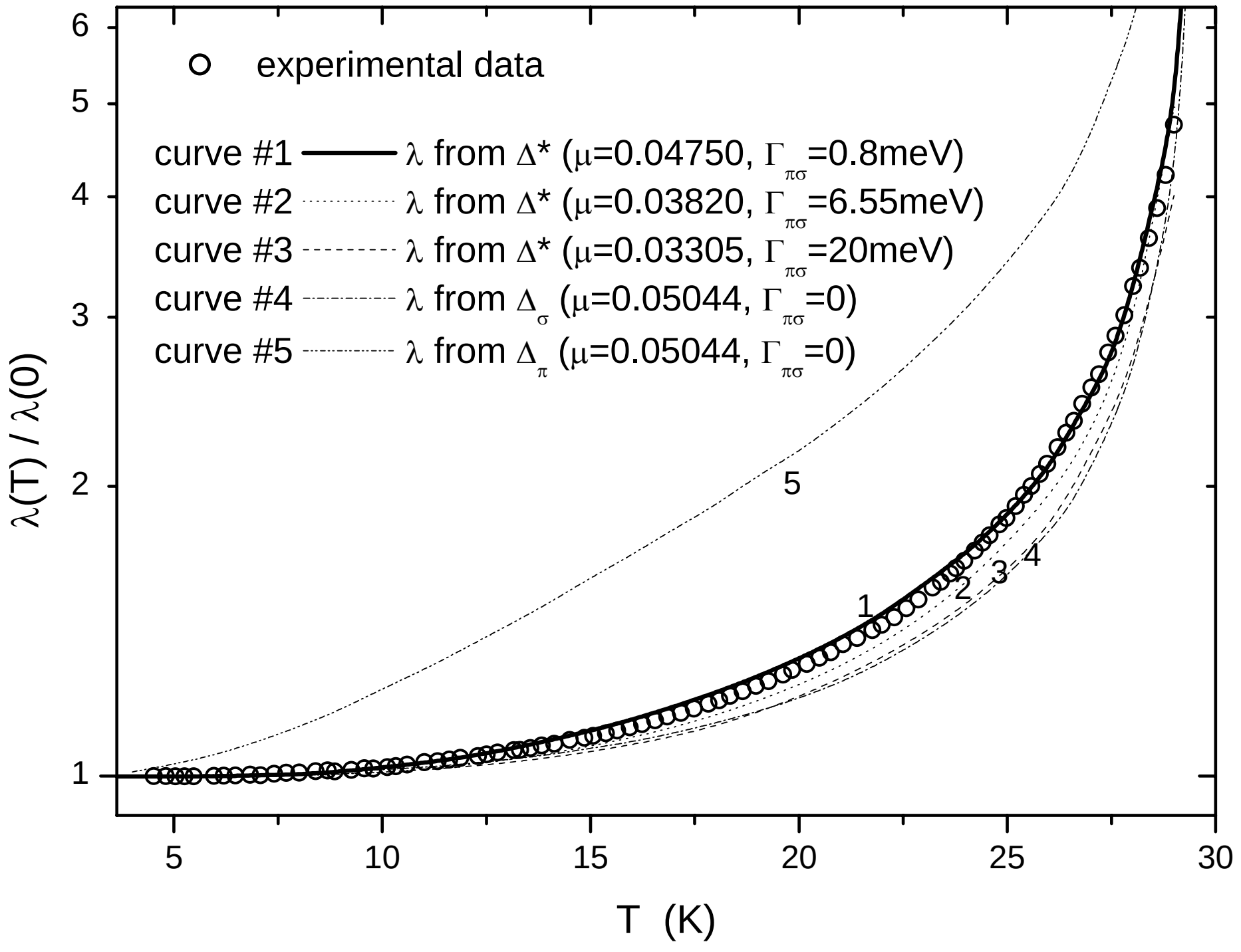

**Figure 7**

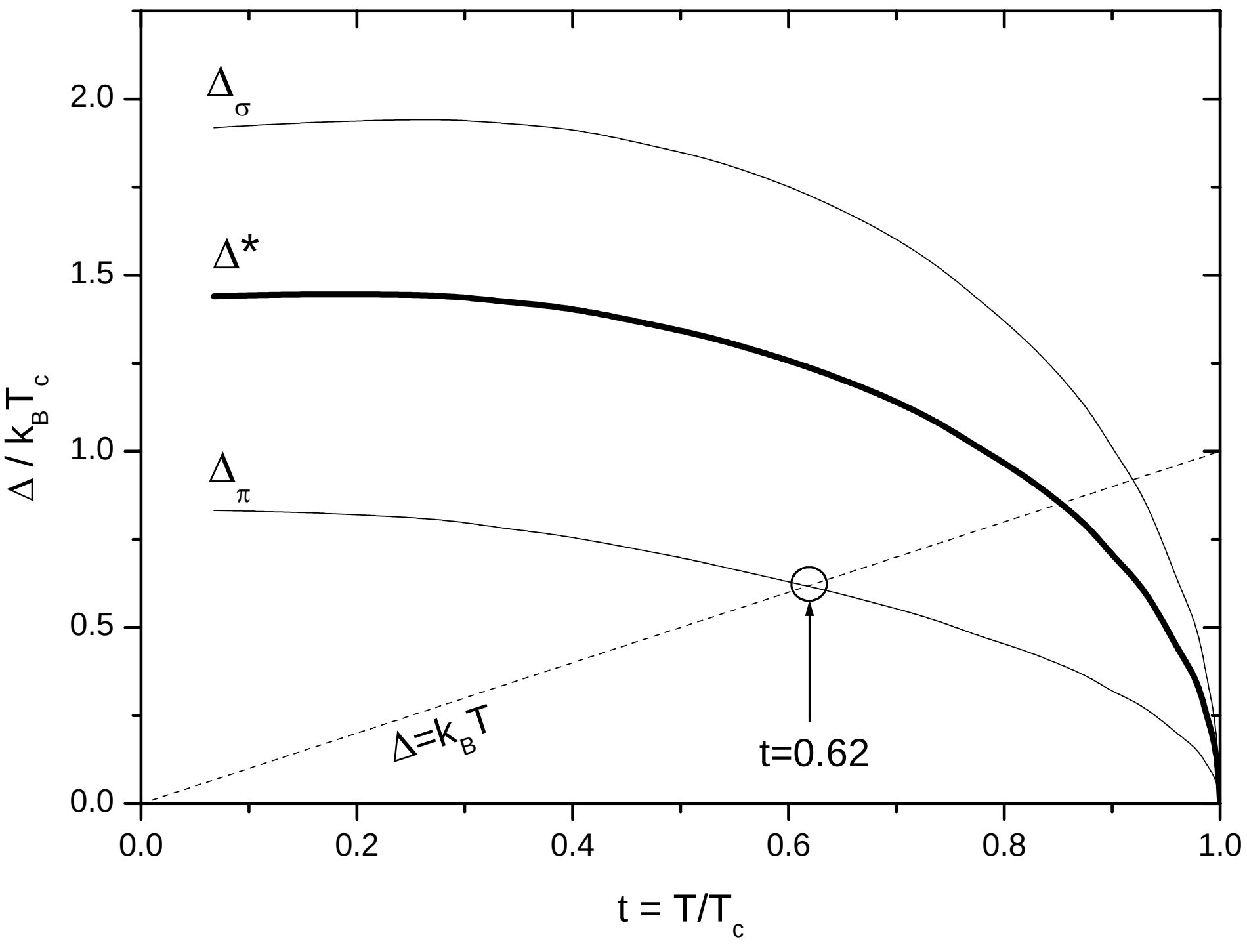

**Figure 8**

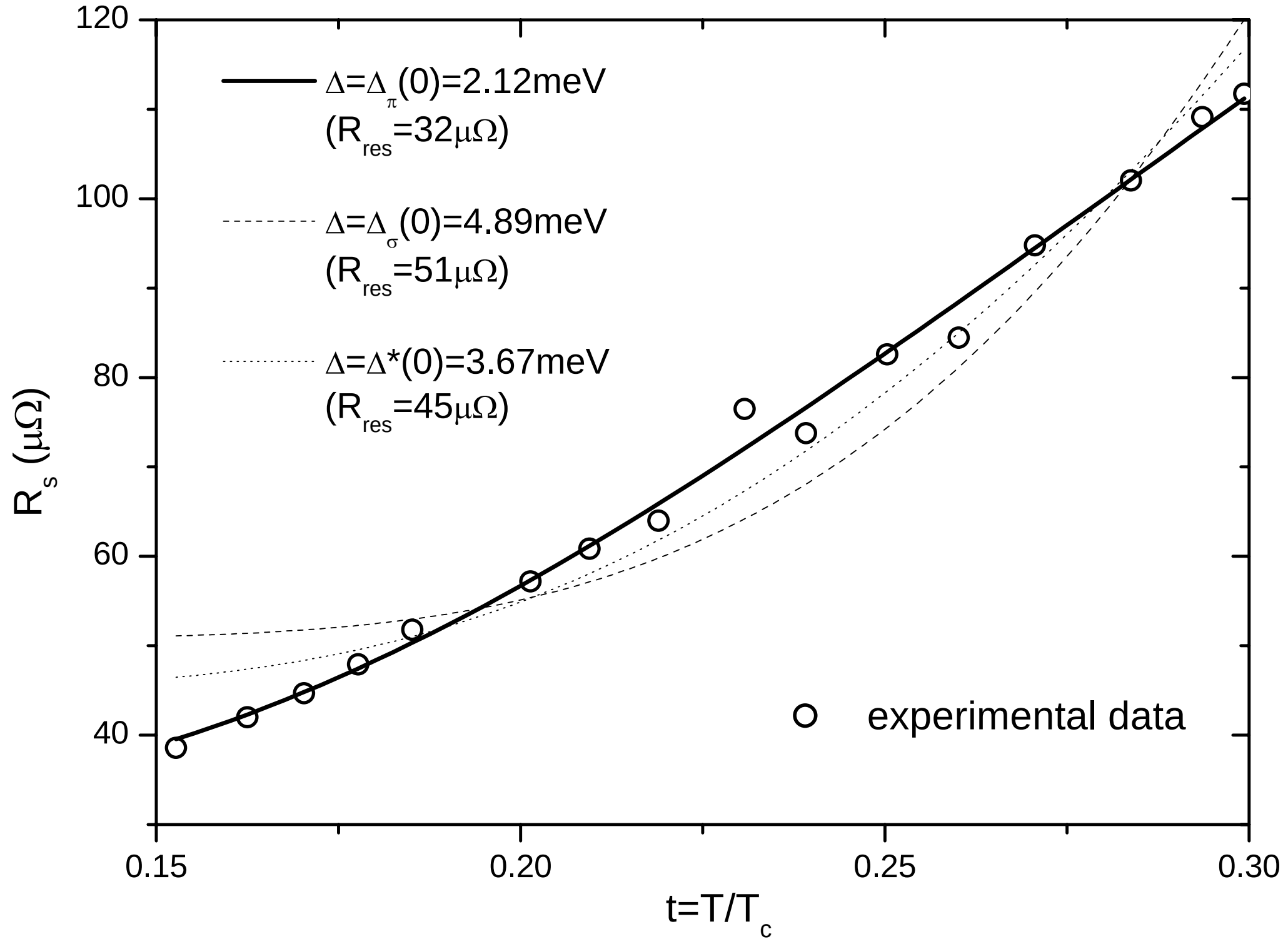

**Figure 9**

Golubov, A. Brinkman, O. V. Dolgov, J. Kortus, and O. Jepsen, Phys. Rev. B **66**, 054524 (2002).

[22] An alternative way to calculate the spectral functions is the approach of Choi et al. (Nature **418**, 758 (2002)). By an iterative technique Choi et al. obtain the full crystal momentum k and temperature dependence of the energy gap of $MgB_2$ from first principles. However, since Choi et al. do not give the band-integrated values of their coupling constants, it would be very complicated to use them in our model. The same holds for the spectral functions. The spectral functions and the coupling constants given by Golubov et al. are easier to apply, since they are averaged over the Fermi surface.

[23] M. A. Hein, in Microwave Superconductivity, H. Weinstock and M. Nisenoff eds., NATO Science Series, Kluver Academic Publisher, p.21 (2001).